\documentclass[12pt
]{article}

\usepackage{natbib}
\usepackage{hyphenat}
\usepackage{hyperref}

\newcommand{\tit}[1]{\textit{#1}}

\begin{document}

\title{Unobservable entities in QBism and phenomenology}

\author{Jacques Pienaar \\ {\footnotesize QBism group, University of Massachusetts, Boston, USA.} \\ {\footnotesize jacques.pienaar@umb.edu}}

\date{\today}

\maketitle

\section{Abstract}
The interpretation of quantum theory known as QBism argues that many elements of the formalism have a subjective interpretation. At the same time, QBism claims to be a broadly realist program. This implies that reality in QBism must be somehow founded upon an agent's subjective experiences (measurement outcomes). To make this idea more precise, we propose to interpret QBism's ``experiences" as synonymous with the concept of ``perceived phenomena" in phenomenology. This suggests an approach to ontology in which objects can only be physically real if they are in principle observable. But what does ``observable" mean? Are atoms, electromagnetic fields, quantum states, or probabilities observable? Here we discuss the different answers to this question given by QBists and phenomenologists, and attempt to reconcile them.

\section{Introduction}
Unlike most other interpretations of quantum mechanics, \tit{QBism}\footnote{See \cite{FS_coherence,FS_greeks,FMS_intro,fuchs_perimeter,fuchs_PR,fuchs_nwb,FAQBism,FS_handbook}.} departs radically from traditional forms of scientific realism by insisting that the quantum state of a system only refers to an agent's subjective beliefs about the likelihood of outcomes of various measurements they might perform on the system. Furthermore, QBism holds that the outcome of any such measurement is an experience of the agent performing the measurement, and so is intrinsically ``personal"\footnote{A more accurate term would be ``indexed to the agent", since QBism does not insist that ``agent" refers to a person \cite{FAQBism}.} to the agent.

Despite its emphasis on the subjective aspects of quantum theory, QBism sees itself as a methodological program that aims towards an ontology of the physically real  \citep{fuchs_PR, fuchs_interview}. To this end QBism seeks to identify those elements of quantum theory which do not admit of a purely subjective interpretation. While these elements would still necessarily be grounded in agents' ``experiences", their explanation would require making some definite claims about ``how the world is" for \textit{all} agents, that is, such elements would point towards the ontological content of quantum theory. QBism thereby acknowledges that an agent's ``experience" does not refer to an isolated consciousness contemplating itself, but includes also some transcendent elements (i.e.\ elements that refer to a ``world" which lies beyond the agent's consciousness) and QBism relies upon this fact to refute the charge of solipsism i.e.\ the claim that only one's own mind is knowable through experience. However, in taking both the subjective and transcendent aspects of ``experience" as equally fundamental, QBism contradicts the traditional view that ontology is something \tit{a priori} objective, and consciousness is something reducible to non-conscious ontological elements \citep{fuchs_nwb, Pienaar2021_QBRQM}. In order to successfully defend its unorthodox vision of ``ontology" against more traditional forms of scientific realism, QBism would benefit from a metaphysical framework capable of accommodating the unique demands that it places upon ``experience".

Framed in this way, there is good reason to hope that phenomenologists might be able to provide this guidance. Like QBists, phenomenologists seek to ground scientific claims about the world in phenomenal experience, and like QBists they reject traditional mind-world dichotomies and seek an ontology grounded in phenomena, which have an \tit{intentional} character that makes them neither wholly immanent to consciousness nor wholly transcendent \cite{Zahavi_basics}. These similarities make phenomenology a natural setting for elaborating a metaphysical thesis underlying QBism, thereby clarifying and fortifying QBism's ontological position.

The present article aims to investigate how QBism might benefit from phenomenology in articulating its response to a particularly vexing question: which of the ``unobservable" entities that appear in scientific theories, such as atoms, quarks, and quantum states, have a claim to physical existence, i.e.\ to be regarded as worldly objects? 

QBism's stance on this issue is presently unclear. QBists often tacitly assume that ``systems" exist which have a mostly objective character, in the sense that the same system can be addressed by multiple agents; these systems include not only straightforwardly observable objects like tables and trees, but also scientific objects like atoms and fundamental particles whose observability may be questioned. In the latter case, the ``subjective" component of these systems is relegated to the quantum state that is assigned to them, which characterizes the agent's expectations for the outcomes of measurements that could be performed on the given object. 

This tacit splitting up of scientific phenomena into an objective ``system" furnished with a subjective ``quantum state" seems problematic on at least two counts. Firstly, particularly in quantum mechanics, there seems to be a certain arbitrariness in what constitutes a ``system" as distinct from a ``property of a system". For example, at low energies, one can take a single electron to be a system, insofar as it remains identifiable as distinct from other electrons. At high energies, this distinction becomes impossible due to particle creation and annihilation processes, and it is more appropriate to regard an electron as an outcome of a measurement (say, checking the ``number of quanta") of a more complicated system that is a ``quantum field". Secondly, both the systems and their quantum states appear to have similar roles in quantum theory, in that both are ``unobservable" entities which serve to co-ordinate and constrain the theoretical tasks of prediction and explanation of observable experimental data. Given this similarity, what reasons can the QBist have for treating ``systems" as being more objective than the ``quantum states" assigned to them? Since some phenomenologists have explicitly discussed the metaphysical status of ``unobservable" scientific entities, QBists might hope to find answers by drawing on the phenomenological literature.

This article is structured loosely as follows: we begin by clarifying what is meant by ``unobservable scientific entities". We then discuss a selection of the phenomenological literature on this topic, which will lead us to divide the question into two key components: (i) whether scientific entities are unobservable in principle or only in practice, and (ii) whether belief in the physical existence of in principle unobservable entities can nevertheless be justified. We then propose a loose connection between QBism's notion of ``experience" and the phenomenologist's notion of ``perception", and we use this connection to explicate QBism's answers to the above key questions. Finally, we identify two aspects of the QBist's replies that pose a challenge to phenomenologists, and we outline some avenues for reconciliation.

\section{Unobservable entities}
What is an ``unobservable scientific entity"? Many objects studied by physicists seem to be ``unobservable", in the sense of not being directly perceptible in the same manner as ordinary everyday objects. We can see the legs of a table just by looking at it or touching it, but we cannot ``see the atoms" in the table by any similar manner. We physicists are nevertheless disposed to claim that we are justified in believing in the atoms' existence, i.e.\ that although we can't see them, we know they must ``be" there as physically real entities. The justification for this claim seems to require something more than what is ordinarily perceptible -- but what is this ``something more"?

To begin with, we must not commit the fallacy of thinking that abstract or theoretical objects are unobservable \tit{per se}. Phenomenologists emphasize that every object given in perception is given within a certain context or ``horizon", against which it takes on certain significance. This significance might include the object's being a concrete instance of some theoretical abstraction, or embodying some formal mathematical model. For example, a set of three billiard balls may take on the significance of being a concrete instance of the number ``three", just in case one is interested in counting them; alternatively they may be seen as concrete instances of perfectly rigid and uniform spheres moving on a two-dimensional frictionless surface, just in case one is interested in predicting their dynamics on the billiard table with a low degree of precision. Even for less overtly analytical tasks, like recognizing that the objects are ``billiard balls", makes reference to a context in which it acquires its abstract meaning, in this case the context is that of games involving cue-sticks and balls. To the extent that ideal theoretical entities like perfect spheres can be concretized in perceived objects, we can consider them ``observable" despite their conceptual or abstract character; paraphrasing Merleau-Ponty, ``the matter of perception is pregnant with its form" \citep{merleauponty_PriP, merleauponty_PP}. 

Significantly, certain scientific entities including atoms, quarks, and electromagnetic fields, do not seem to ever appear concretely as perceived objects. We know of them only indirectly, by inference from the behaviour of certain directly perceptible objects that serve to signify their presence. For example, we can use a scattering of iron filings to infer the presence and partial structure of an electromagnetic field, and the readings of a scanning-tunneling microscope to infer the structure of atoms on a surface. In short, the phenomenal evidence for the existence of such entities typically comes in the form of graphs, numerical readings, motions of mechanical sensors, etc, which serve to indicate these entities but do not present the measured objects ``bodily" in perception; thus the entities themselves are said to be ``unobservable".

\section{Subjectivity and objectivity \label{sec:SubjectObject}}
The question naturally arises as to whether this unobservability is a matter of principle or of practice. 

According to QBism, the answer depends on the nature of the unobservable entity. QBists implicitly make a distinction between two types of unobservable entity: those that have the character of being ``system-like", versus those that have the character of being ``probability-like". The former kind are entities that are regarded by the agent as external objects to which measurements can be specifically addressed; they are uniquely identifiable by some label and a set of characteristic properties that define their type (in the case of an electron, these would be: a point-like spatial extent, a mass of $9.109\times10^{-31}$kg, quantum spin of one half, \tit{etc}). Such entities may include planets, chairs, fields, atoms, and fundamental particles. The second kind of entity are those which are explicable as ``catalogs of probability assignments". The most basic of these is just a probability distribution over a complete set of mutually exclusive experiences the agent could have. More complicated examples are quantum states, Hamiltonian operators, and POVMs, which the QBist interprets as being ultimately equivalent to sets of probability assignments (or in some cases, conditional probability assignments) about possible experiences that the agent could have, specifically, experiences of the outcomes of certain measurements (i.e.\ elaborate actions) they could perform.

For a QBist, the system-like entities have a more ``objective" character, while probability-like entities have a more ``subjective" character. This terminology is unfortunate, because these terms are a relic of the Cartesian split between a mind-bearing ``subject" placed in categorical opposition to an independent world of ``objects". A close reading of QBism suggests that they reject such an absolute division. 

To see why, let us set aside ontological questions for the moment and consider QBism strictly within the context of decision theory. In this restricted context, QBism asserts that quantum theory is a ``tool of reasoning", that is, a set of norms or best practices that guide the decisions of agents. Implicit in this formulation is the idea that the users of quantum theory belong to a special class of agents, i.e.\ not only must they be capable of reasoning explicitly using symbolic language, but they must have specific concerns for which the usage of quantum theory confers an advantage, such as building a quantum computer, or designing high-precision particle detectors. To the extent that QBism distinguishes system-like entities from probability-like entities, this distinction is relative to the particular ``set of possible agents" who are implicated as the users of the specific decision theory (in this case quantum theory). Given a specification of the community of possible agents, one may call ``subjective" those quantities or entities which may vary from one agent to another within the community. Conversely, those entities which preserve their form across the whole community may be called ``objective", provided we understand this word as meaning ``invariant between subjects" and not ``independent of subjects" (which is how physicists usually understand this word).

Seen in this way, it also becomes clear that the traditional terms ``subjective" and ``objective" are inadequate to capture the full meaning of the distinction. Firstly, even the most ``subjective" elements -- the experiences of the individual agents in all their richness and idiosyncracy -- are only conceivable within the broader ``objective" norms that were posited at the beginning to define ``the set of possible agents". The mere fact that we have taken up a specific decision theory (eg.\ quantum theory) which is only relevant to a specific class of agents means that we have imposed a set of implicit ``objective" features that are shared by all agents, and shape their experiences. Secondly, even the most ``objective" elements -- like the abstract rules of quantum theory themselves -- have their origin and grounding in the ``subjective" raw material that comprises the history of observations by individual agents within the community, subsequently communicated, refined, modified to accommodate new experiments, and ultimately processed by that community until invariance was achieved (at which point the rules become `laws' and are disseminated in textbooks and institutional curricula). In summary: variance can only be defined with respect to the invariant, and new invariants can only be extracted from a substrate of initial variance; there is always a seed of the objective in the subjective and vice-versa.

So long as QBism remains only a kind of decision theory, these subtle matters need not trouble it. For a decision theory does not require the class of possible agents to be explicitly named and characterized, in the same way that you do not need to be explicitly aware that you have opposable thumbs in order to use a hammer -- you simply pick it up and start using it. This is why QBists are constantly frustrated by criticisms to the effect that they do not bother to carefully define ``what is an agent". Their reply is the same as the decision theorists' reply: there's no need for such theorizing, just try to use the ``hammer": if you can, you are an agent of the relevant sort; otherwise, you are not. Thus, QBists often use the word ``subjective" to characterize various parts of quantum theory (quantum states, Hamiltonians and so on) without really bothering to give it a precise definition, aside from remarking that these things are subjective by virtue of being able to differ markedly from one agent to the next. As such, the ``subject" vs ``object" distinction persists in QBist language, despite the fact that their own metaphysical commitments ultimately push them to go beyond this naïve dichotomy.

This tension becomes impossible to ignore when QBism aims beyond mere decision theory and tries to articulate an ontology. For, if QBism follows the example of phenomenology, it must be acknowledged that the ``agent" which appears at the decision theoretic level must not be separable from their ``external world" at the fundamental level. QBism must seek a notion of ``pre-reflective" ontological being which is rich enough to give a historical account of the evolution of the agent-world split. In the process, we must lose any sharp distinction between ``objective" and ``subjective", these being regarded as idealized extremes on a gradient of degrees, and the classification of phenomena as being ``more objective" versus ``more subjective" will then be seen as a contingent process which is never completed, but is constantly evolving and being revised as the emergent ``self" contends with the emergent ``world" through the phenomena in which they are mutually bound. Incidentally, this is also why it is impossible for QBists to talk about the transformation or mutation of the agent without also making ontological commitments that go beyond decision theory, as in eg.\ \citet{pienaar_2020}.

This is relevant because in QBism some unobservable entities have a ``more objective" character, and others a ``more subjective" character, and this has bearing on whether they can be observable in principle or not\footnote{Note that in light of the preceding discussion, the `observability in principle' of something must be understood as being relative to a particular set of possible agents.}.
System-like entities are ``objective" in the following sense: an agent can experience that there is a system before them upon which they can do measurements; moreover it is unproblematic to assume that this system can be identified by a unique label over time and can be shared with other agents who can be said to perform measurements on the ``same" system. On the other hand, the probability-like entities have a ``subjective" character in the sense that they represent the probabilities (i.e.\ quantified degrees of belief) of the agent, and therefore neither stand before an agent like an object, nor can be simply taken as shared between agents\footnote{Note that the agreement of two agents' beliefs in one or more aspects would not imply that the agents hold the same belief, any more than the identical appearances of two apples would imply that they are the same apple.}. In particular, by virtue of the personal nature of probability-like entities (i.e.\ their variability across agents), QBists insist that they cannot be regarded as physical entities, i.e.\ entities which exist as things in the world external to the agent; therefore they are not even in principle observable in the sense of being perceptible as physical things. 

We pause to remark that this view about the character of probability-like entities is a direct consequence of QBism's commitment to a subjective Bayesian interpretation of probabilities \citep{BernardoSmith1994}. According to this view, there is no such thing as a ``wrong" prior probability assignment. In the present context, we may think of this as providing QBism with a constructive principle that says something about how we ought to define the ``class of possible agents", namely, we should define it to be as permissive as possible, by including all agents whose beliefs can represented by some valid probability function.

\section{The Prolongation Thesis}
In contrast to probability-like entities, there seems to be a precedent in QBism for regarding system-like entities, including atoms, electromagnetic fields, and the like, as being in principle observable. The crucial ingredient is the QBist principle which states that measuring instruments are to be regarded as ``prolongations" of the agent's senses.

It is worth explaining in more detail what the QBists mean by this. As recalled in \citet{fuchs_nwb}, the principle originates from a remark by Wolfgang Pauli in a letter to Niels Bohr in 1955, where Pauli suggests that ``\dots it is allowed to consider the instruments of observation as a kind of prolongation of the sense organs of the observer\dots". Pauli later retreated from this idea, but QBism adopted it as a core principle. QBists saw this as a necessary move in order to resolve the ``Wigner's Friend" paradox. We need not review the details of this debate here (see \citet{fuchs_perimeter, fuchs_interview, wf_fellow}); it suffices to note that QBism criticizes the Copenhagen tradition (as variously propounded by Bohr, Heisenberg, and Pauli) as being unable to address this paradox, because of its refusal to allow measurement outcomes to be ``personal" to the agent, which is implied if one takes the instruments to be prolongations of the agent's senses. As \citet{fuchs_nwb} puts it:

\begin{quote}
QBism [\dots] takes [Pauli's idea] deadly seriously and runs it to its logical conclusion. This is why QBists opt to say that the outcome of a quantum measurement is a personal experience for the agent gambling upon it. Whereas Bohr always had his classically describable measuring devices mediating between the registration of a measurement’s outcome and the individual agent’s experience, for QBism the outcome just is the experience. --- \citet{fuchs_nwb}, p21.
\end{quote}

We will encapsulate this idea in the following statement, which we call the ``Prolongation Thesis":

\begin{quote}
Prolongation Thesis: Those measuring instruments which the agent regards as being the source of their experiences (i.e.\ of their measurement outcomes) are to be regarded as prolongations of the agent's body, and thus having the same metaphysical status as the bodily sense organs of the agent.
\end{quote}

It is important to note that QBists do not intend this to be a claim about ``measuring instruments" as defined by some external criterion. Rather, in asserting that measuring instruments are prolongations of the agent, QBists are proposing that this is \tit{definitive} of what a ``measuring instrument" means for the agent.

Specifically, the QBist is here implying that to call something a ``measuring instrument'' is to assert that its presence to oneself and its usage is comparable to that of any other of one's given bodily sense organs. In particular, the experience of perceiving an object through the instrument must be no less direct, immediate, effortless, and unmediated as the experience of perceiving an object via one's familiar sense perceptions. Thus an instrument whose outcomes are inferred by explicit theoretical work, tedious adjustments, or any other cognitive processes exceeding those normally employed in the usage of one's ``natural" bodily functions, must strictly speaking not be a measuring instrument on the QBist definition.

This terminology is admittedly awkward. It implies that QBists do not accept the meaning given to ``measuring instruments" in common scientific parlance, where things such as microscopes and telescopes are taken to be \tit{de jure} measuring instruments, irrespective of the operator's skill and experience while using them. Given the fact that most scientists likely do not feel the same facility with their instruments as they do with their own hand or eyes, it appears that most things the physics community considers to be measuring instruments do not meet the QBist's definition.

One way to navigate this overloading of terminology is to follow \cite{pienaar_2020} and make a principled distinction between apparatuses external to the agent that satisfy the conventional desiderata of scientific instruments, as opposed to those apparatuses with which the agent is so fluent that they may be considered equivalent to additional sense organs. Here we endorse this distinction by referring to ``extracorporeal" versus ``intracorporeal" measuring instruments, respectively.

Returning to the question of ``unobservable entities", we see that the observability of the system-like entities depends on whether the initially \tit{extra}corporeal instruments that signify the system's presence can become \tit{intra}corporeal after sufficient training and practice; that is, whether the instrument used to detect the system could ever in principle be taken as being on par with the agent's own sense organs. If it could, then we would have to conclude per the Prolongation Thesis that the systems which can be made detectable to the agent through their intracorporeal instruments are in principle ``observable", in the same manner that any other physical thing is observable to the agent.

We must therefore ask: is it conceivably possible for an initially extracorporeal instrument to become intracorporeal? As a possible counterexample, consider that the detection of the Higgs Boson by the Large Hadron Collider involves an apparatus that is not only incredibly huge, but involves a collaboration of thousands of scientists, engineers, and other specialists, all working in concert to collect, prepare, and analyze the data. Although it is certainly tempting to claim that such an instrument could never conceivably be ``intracorporeal" for any agent, there are some reasons to hesitate in making this judgement. 

Firstly, following decision theory, QBism by no means restricts ``agents" to be individual people; in fact QBists have at times explicitly endorsed the idea that groups of people could be regarded as agents in the relevant sense. If one could make this idea explicit, say, by positing what constitutes an ``experience" for a collective, then the human operators of the LHC might well be considered a single agent, and (exotic as it sounds) the LHC considered as one of their ``sense organs". 

Secondly, even if the LHC specifically cannot be conceivably treated as intracorporeal to an agent (even a collective), we cannot easily exclude the possibility that one could build a more compact and automated device for detecting Higgs Bosons, which might itself be amenable to such fluent usage as to become intracorporeal to the user.

Since neither of these possibilities can at this stage be readily negated or affirmed -- i.e.\ because we lack both a systematic and rigorous account of what it could mean for a collective to be an ``agent", and we lack any arguments for or against the plausibility of a hand-held Higgs detector -- we must leave it open as to whether Higgs Bosons are in principle observable or not, according to QBism's Prolongation Thesis.

Fortunately, for present purposes, it is enough to note that at least some unobservable system-like entities are detectable by instruments which could plausibly become intracorporeal for an agent, thereby rendering such entities observable. A case in point is the example of a scientist using a scanning tunneling microscope (STM) to measure atoms. In \tit{Meeting the Universe Halfway} \citep{barad}, Karen Barad recounts physicist Don Eigler's dramatic live demonstration of the usage of the STM to the audience of National Public Radio's 1996 broadcast of the Morning Edition. Barad leaves little doubt of the visceral nature of the experience for both the operator of the device and the listeners:

\begin{quote}
With a few clicks of the computer mouse, Eigler maneuvers the STM tip so close to a gadolinium atom sitting on the surface of a piece of niobium
that it begins to bond with the gadolinium atom. He moves the tip sideways, pulling the gadolinium atom across the niobium surface to a new location,
and then pulls the tip back, releasing the atom. The listening audience is treated to a sonic display of the single-atom manipulation, courtesy of Eigler's clever connection of the STM to a stereo that converts the strength of the ``tunneling current" (used to sense the presence of an individual atom) to an audible tone
[\dots]
The proof is in the hearing. During the sideways tug of the gadolinium atom across the niobium surface, the audience hears distinct ``thunks" as the atom is pulled across the unit cell structure formed by the spaces between the niobium atoms on the surface: that is, one can hear the atom
being moved. --- \citet{barad}, p355.
\end{quote}

This account suggests that it is possible to create an interface to the STM which seamlessly connects a user's bodily motions to sensory feedback from the measured atoms, giving them an especially intimate experience of the measurement. Perhaps with enough practice, the user might come to feel as if they really are hearing or touching atoms as they move them around. However, this is merely a plausibility argument: we do not claim to have established that such an interface would be sufficient to give the atoms a ``bodily presence" in perception.

In fact, this argument poses the first challenge to phenomenologists: can one make sense of the QBists Prolongation Thesis in a phenomenological context? This seems likely to be a contentious question in phenomenology. On one hand, in defending the in principle unobservability of atoms and the like, \citet{wiltsche_2012} explicitly refutes the ``argument from mutation", which states that advanced scientific instruments may become organic parts of an observer's sensory apparatus through evolution or prosthesis -- essentially the argument we have just presented in support of the Prolongation Thesis. On the other hand, phenomenologists including \cite{merleauponty_invis} and \cite{Todes_BW} take the body to be the mutable and changeable ground of perception: the difference between bodily and non-bodily objects is on their account a difference of degree but not of kind. This seems to leave open the door for a revival of a (perhaps more sophisticated version of) the argument from mutation. But here is not the place to speculate: we leave the issue open to debate.

\section{Justifications for physical existence}

In light of the foregoing, there may yet be some entities which are not observable even in principle. This provokes a second question concerning such entities, namely: is their in principle unobservability due to their having a purely conceptual character (like probabilities), or can these entities be present as things in the world like tables and trees, in which case their unobservability is due to them being ordinarily invisible (like X-rays or viruses)? In the first case such entities would be seen as glorified ``book-keeping devices" (\tit{Hilfsmittel} as Husserl would put it) that we use for scientific reasoning and prediction, and would not be considered ``scientifically real". In the second case we would have to provide independent arguments (i.e.\ besides the possibility of direct observation) that could establish the existence of these entities as physical things within the world, about which we could be justified in saying that they are ``really present before us", even though we could not ever directly observe them. We may frame this as the question of whether we can have \tit{epistemic justification} for believing that something exists as a thing in the world, given that it is in principle unobservable.

In the phenomenology literature, \citet{wiltsche_2012} argued that Husserl endorses a particular criterion of epistemic justification. In a subsequent critique of this criterion, Berghofer named it the  ``Originality Thesis of Justification" (OTJ) and gave it this definition:

\begin{quote}
Epistemic justification is limited to what can be originally given in the sense that if X cannot be given in an \textit{originary presentive intuition}\footnote{This technical term -- givenness in originary presentive intuition -- is here understood as referring to X's being an object of the agent's sense perception}, then one cannot be justified in believing that X exists or obtains.  --- \citet{berghofer_2018}, p4.
\end{quote}

Adopting this criterion implies that if an entity cannot be observed in principle, then one cannot be justified in believing that it is a really existing thing in the world\footnote{Note that one is not forbidden to \tit{believe} that such an entity physically exists -- just that one's belief cannot be justified on `epistemic' grounds. The question of what other grounds there might be for believing in something is left open to speculation.}. However, \citet{berghofer_2018} points out that Husserl's position is more ambiguous, and that phenomenology does have the resources to accommodate a criterion of justification that would allow for justifiably believing in the existence of unobservable entities. 
Berghofer's strategy is to appeal to the phenomenological concept of a \tit{horizon}: when something is given originarily in perception, it is not simply given as a dead and lifeless fact, but as a carrier of associated further possible experiences which are \tit{co-given} with it. Some of what is co-given are things that can themselves in principle be given originarily, for instance, when one perceives a house from the street one also perceives that it has an interior, and this co-given interior can become originarily given by entering the house and perceiving it.

Berghofer argues that one can also consider co-given entities which cannot be given originarily in perception even in principle, but whose existence can still be justified through their implications for other things that can be given originarily. For example, although we cannot perceive muons directly, they are co-given in our perceptions of the various forms of observable evidence for them, such as tracks in cloud chambers or detector clicks, which serve to indicate their physical presence. They can thereby shape our expectations about what will happen in future observations, which can subsequently be fulfilled in originary perceptions. Thus, following Berghofer, we can be justified in believing in the physical existence of muons despite their in principle unobservability. Berghofer names this alternative the ``Criterion of Justification" (CoJ):

\begin{quote}
Justification with respect to content C or object O is only possible if C/O can either be originally given or if C/O can be reasonably associated with expectations that can be verified by originary presentive intuitions.  --- \citet{berghofer_2018}, p11.
\end{quote}

We note that Berghofer does not claim that this criterion is sufficient for justification: he merely opens the door for such justification to be possible. Nevertheless, this criterion is more permissive than the QBist would like, because it seems to permit some things to exist as physical things which the QBist would very much like to exclude from that category. 

As we have mentioned, QBism adheres to a subjective Bayesian interpretation of probability, and considers quantum states (among other parts of the quantum formalism) to be equivalent to probability assignments, which in turn have a strictly conceptual character: they are symbols used by agents for the purposes of reasoning about their experiences. Significantly, the agent does not experience probabilities as things that stand before them in the world: though they might `exist' to the agent in a certain general sense, they do not exist \tit{as physical things} (according to QBism). The fact that they may have \textit{correlates} in things which are physical -- such as patterns in the firing of the agent's neurons, or bets the agent subsequently places in the casino -- does not refute the QBist's fundamental position that probabilities themselves are subjective in the highest degree, and thus are not `things' like rocks or trees.

The OTJ could provide QBism with the resources it needs to defend this position. One only has to argue that probabilities are not themselves directly observable in principle, that is, while they can be \tit{inferred} from observable phenomena (neuronal firings and bets at the casino), there is not and can never be any instrument that would let one agent directly `see' (or hear, touch, etc) another agent's beliefs and hence their probability assignments. To the extent that such an argument is defensible, the OTJ would allow us to conclude, following de Finetti, that ``probability does not exist" \citep{deFinetti1990}. By extension, this would justify the claim that quantum states are not physical things, as the QBist desires.

By contrast, Berghofer's CoJ is no help to the QBist; since probabilities can indeed be ``reasonably associated with expectations that can be verified by originary presentive intuitions", like neuronal firings and casino bets, CoJ alone is powerless to prevent us from declaring probabilities (and quantum states) to physically exist in a manner not fundamentally different to rocks and trees, forcing the QBist to cast about for independent arguments to exclude this possibility.

Another case in a similar vein is the possible existence of so-called ``hidden variables", which cannot be directly observed in principle, but may be invoked in theoretical models as supplementary properties of physical systems that serve to determine the outcomes of future measurements on them. No-go theorems such as Bell's theorem establish that these models must have extremely counter-intuitive behaviour (such as exhibiting faster-than-light transfer of information), and this is one of the many reasons why QBism rejects the existence of such hidden variables. Indeed, the presumed existence of hidden variables would render QBism irrelevant as an interpretation. Here again, OTJ would prove a useful tool for the QBist: since hidden variables are unobservable in principle, OTJ implies that we cannot be justified in believing that they exist as real properties of physical things. This conclusion vindicates QBism's rejection of hidden variables. By contrast, CoJ alone provides no grounds for such a rejection, since proponents of hidden variable models can argue that they do have observable consequences. 

In conclusion, if QBism is to draw upon phenomenology to address the problem of the existence of unobservable entities, it must do so in a peculiarly opportunistic fashion: on one hand it must side with \cite{berghofer_2018} against \cite{wiltsche_2012} in considering OTJ to be too restrictive, because it excludes things such as atoms, electromagnetic fields, etc, from the possibility of existence; on the other hand, ironically, it is precisely OTJ's restrictiveness that would permit QBism to exclude probabilities from being physically real things. To reconcile this, we propose that QBism endorse a slightly more permissive version of OTJ that expands the scope of what is in principle observable through the Prolongation Hypothesis. This `prolongated OTJ' would be permissive enough to escape many of Berghofer's criticisms of OTJ, while still being restrictive enough to keep the door closed to the physical existence of probabilities and hidden variables.

\section{Husserl and subjective probability}

If QBism does adopt the OTJ as we suggest, then probabilities (and hence all probability-like entities, including quantum states) must be excluded from the domain of worldly existence. Probabilities must rather be thought of as having a conceptual or symbolic character, manifesting themselves not as features of perceived things, but as features of the horizons of perceived things \cite{Tremblaye:2020}. When something is given in perception, a horizon of possible further perceptions is co-given with it, and we may perhaps think of probabilities (hence also quantum states) as modalities or ``weights" of these possibilities within the horizon, corresponding somehow to a sense of ``how likely" each of them is.

This gives rise to our second challenge to phenomenologists: can they provide an interpretation of probabilities that can accommodate QBism's subjective interpretation of probabilities, quantum states and other probability-like entities? There are two reasons why this poses a challenge, one historical and one interpretational.

Historically, after Husserl's incomplete and somewhat confusing treatment of the phenomenology of probability, whose most mature expression is found in his lectures \tit{Logik und allgemeine Wissenschaftstheorie} \citep{husserl_logik}, Husserl's followers have displayed surprisingly little interest in the topic of probability. Merleau-Ponty, Sartre, Heidegger, and other famous phenomenologists barely mention probabilities, except perhaps very obliquely in their discussions of signs and symbols. An exception to this trend is \citet{lobo}, who revisits Husserl's phenomenology of probability and offers some hypotheses as to why it has been neglected by later scholars. Lobo's particular interest is to revive a program, initiated by Husserl and subsequently developed by Gian-Carlo Rota, whose ambitious aim is to reform classical logic to ground it in probability theory, which Husserl came to regard as being the more fundamental formal structure. Our present aim concerns a different program, which is to provide a phenomenological analysis of probability, for which our most recent primary reference still appears to be Husserl's incomplete sketch from over a century ago \citep{husserl_logik}. 

The interpretational challenge presented there is Husserl's ambiguous stance with regards to the question of whether probabilities are ``objective" or ``subjective". Given our earlier discussion of the meaning of these words, it may be anticipated that Husserl, like QBism, does not take these words at face value. Indeed, in what follows we shall argue that Husserl's view of the meaning of probability is compatible with that of QBism, provided that what Husserl meant by the ``objective" aspects of probability are taken to correspond to what the QBists mean by the ``normative" aspects of probability. As we will see, this connection has the added benefit of deepening and enriching the QBist's notion of ``normative" and connecting it to contemporary philosophy of normativity along the lines of Robert Brandom's work on inferentialism \citep{brandom}.

We begin by reviewing the essential points of Husserl's analysis of probability. \citet{husserl_logik} makes three key points about how to interpret probabilities in phenomenology. We can loosely think of the first two points as addressing the ``subjective" aspect of probability, while the third point addresses their ``objective" aspect. 

Husserl's first point is that probabilities, at their most basic level, are not themselves objects of consciousness, but are rather modes of intentionality, i.e.\ they are about something -- an event or occurrence -- which is not itself a probability. More specifically, they are qualities of judgements that one makes about things, and the quality of something being ``probable" is not subservient or reducible to any mixture of absolute certainty and uncertainty, but rather stands on its own as an independent sentiment, existing separately alongside related notions like necessity and questionableness in the ``sphere of judgements". Each of them is a distinct phenomenal quality that ``has its own source", as Husserl explains (our emphasis):

\begin{quote}
Subjectively speaking, we make an affirmative judgment about possibility or probability when we say, ``It is possible that S is P", ``It is probable, questionable, that S is P". The concepts ``possible", ``questionable", ``probable" are thereby not specifically apophantic categories, if we understand apophansis as judgment, as supposed truth. \tit{They just arise from their own sources}. Supposed truth in certain ways stands apart in a parallel manner: supposed possibility, probability, questionableness.  --- \citet{husserl_logik}, p249-250.
\end{quote}

Husserl's second key point has to do with the mysterious ``sources" of probability judgements. As Marian Evans (as George Eliot) wryly remarked in her famous novel \tit{Middlemarch}, judgements often seem to depend upon ``a sense of likelihood, situated perhaps in the pit of the stomach or in the pineal gland". But Husserl cautions that there is much more to probability assignments than the accompanied feelings, whether emotional or physiological. Husserl was cognizant of the fact that any successful phenomenological definition of probability would have to be able to account for the existence of strict rules governing its structure, namely the rules of the probability calculus. These rules appear to serve as good advice to anyone, regardless of their particular feelings, and this robustness against variation from one individual to the next must also be accounted for by the phenomenal character of probabilities.

To achieve this, Husserl argues that probabilities have their source in a very special type of ``possibility", which must be distinguished from other types. Specifically, probabilities are those possibilities that we may call \tit{presumptions}, which have the character of being weighted or ``loaded" by the contextual presence of other facts and things that either ``speak" for or against it. In Husserl's words (our emphasis),

\begin{quote}
Empty possibilities are possibilities in whose favor nothing argues. They are mere figments of the imagination or imaginable things. A well-founded ``possibility", or something presumable with a reason for it, is something entirely different. In all presuming, something figures to the consciousness 
as presumable, i.e.\, just as something argues in favor of this or that. [\dots] But, of course, it is a presumption and not a belief. It is not expressed as a judgment, as an affirmative positing. In our eyes it is not yet truth, but only ``probability". Probability -- originally and understood outside of any relationship –- is the same as the possibility occurring here, and again the same presumability. \tit{And, for its validity, every probability in this sense requires a reason for being probable}. --- \citet{husserl_logik}, p252.
\end{quote}

Husserl's third point is that probabilities can thereby be compared and placed within an ordered structure of gradations. Since reasons can ``speak" more or less strongly in favour of presumptions, it is possible to compare presumptions according to their relative gradations of how strongly they are ``spoken for". In discussing this comparison between probabilistic weights, the word ``objective" appears (our emphasis):

\begin{quote}
Now, however, presumptions come into relationships. You immediately understand what is meant if I point to the familiar way of speaking, ``This argues in favor of it, that other thing also argues in favor of it, and still more is in its favor". I am speaking from a subjective perspective of stronger and weaker presuming. We say that the presumption grows stronger with the number of reasons for presumability. However, we also say that the more there is that argues in favor of something, namely, in favor of the fact that something exists, the greater is its probability; the more that argues against it, the more the probability diminishes. This is an objective way of speaking [\dots] \tit{[A] unique originary relation of precedence and gradation belongs to the presuming-meanings, and accordingly also objectively valid laws of gradation}. In these relationships of gradation what, considered in isolation, was already called probability, is then referred to as mere possibility, and possibilities are weighed against other possibilities [\dots] with respect to the reasons for them or the weight that they carry. --- \citet{husserl_logik}, p252.
\end{quote}

Although it might seem paradoxical that Husserl should emphasize the ``objective" character of probabilities, having earlier emphasized their ``subjective" character, we can unscramble Husserl's omelette of subjectivity and objectivity by appealing to our earlier strategy of translating these concepts into statements about variability and invariance with respect to a given set of possible agents. 

To this end, we note that in Chapter 5 of \citep{husserl_logik} Husserl distinguishes `simple' judgements from `composite' judgements. As Lobo points out (\citet{lobo}, p517) we should therefore correspondingly distinguish ``first-order" from ``second-order" probability statements. First-order statements are those in which probabilities figure in judgements about something non-probabilistic, such as ``I consider it very probable that it will rain", whereas second-order statements take probabilities assignments as their objects, such as ``the probability of rain is 1 minus the probability of not-rain". While first-order statements are evidently always subjective (since they depend on the one making the judgement) second-order statements can capture relations between probability assignments which are independent of the particular values of the assignments themselves, hence are ``objective" in the relevant sense. And if one carefully reads the last passage we quoted from Husserl, it appears that the kinds of statements he cites as ``subjective" and those he cites as ``objective" do indeed fit into the respective categories of first-order and second-order probability statements.

Having thus clarified Husserl's three points about the interpretation of probabilities, let us compare them to QBism. Husserl's first point is evidently quite compatible with QBism, since it asserts that probabilities are (in Husserl's terminology) ``qualities of judgement" and therefore may vary between subjects, and so are ``subjective" in the sense relevant to QBism. 

Husserl's second point differs from QBism in its emphasis: whereas Husserl emphasizes that agents must supply ``reasons" that ``speak in favour" of their probability assignments, QBism ignores the agents' reasons for their assignments, following instead the decision-theoretic line of only demanding that agents commit to purchasing or selling hypothetical lottery tickets by which they stand to gain or lose according to their probability assignments. While not an outright contradiction, this difference in emphasis deserves to be elucidated, and we shall return to it in the next section, after dealing with Husserl's third and final point. 

Husserl's third point might seem problematic because of its reference to some ``objective" aspect of probability. Given our explication of what ``objective" means in this context, however, it should be noted that in QBism the rules of the probability calculus are presumed to be the same for all agents; indeed, this is what QBists mean when they posit that agents strive to be ``rational" and to be ``coherent". Therefore despite maintaining the subjective status of probability assignments, QBists allow that these enter into abstract relations whose form is invariant among the agents, and so we may call it ``objective" in the relevant sense, although QBists prefer to use the word ``normative" to avoid giving the false impression that the equations describe something independent of the agents as a group. 

We may therefore reconcile Husserl's third point with QBism, if we read Husserl's usage of ``objective" as indicating the rules of probability theory in a ``normative" capacity. We can say, along with Husserl, that the rules of the probability calculus are ``second order" constraints on probabilities, and are ``objective" (in Husserl's words) only in the sense that they represent ``normative" rules (in QBism's words) which all rational agents adhere to.

\section{Probability and normativity}

Finally, let us return to the interesting tension regarding Husserl's point about the sources of probability weights. At a glance, there might seem to be an inconsistency here: does a probability weight indicate the relative strength of the ``reasons that speak for it", or does it represent the amount of money at which an agent would be prepared to commit to buying or selling a hypothetical lottery ticket? We now argue that the answer must be: both!

To see that the two views are compatible, note that despite QBism's emphasis on interpreting probabilities in terms of hypothetical bets (an attitude inherited from subjective Bayesian probability theory), it also provides the conceptual infrastructure for any agent to provide ``reasons" for his or her probability assignments, if questioned. Formally, following Bayesian methods for reasoning with probabilities, an agent begins at some point in the past by making a set of ``prior" probability assignments (a Bayesian prior). If the agent is surprised by subsequent experiences, they revise their probability assignments to recover `coherence' among their beliefs and their experience, as represented by certain formal constraints on their probabilities (such as Bayes' rule). Therefore, for a QBist, the ``reasons" behind a probability assignment comprise the agent's prior assignment, a history of their experiences, and the normative rules they employ to keep their probabilities mutually coherent in light of those observations. 

It is worth emphasizing that despite the abstractness of this formulation, it can perfectly well capture what we ordinarily consider to be reasons for probability assignments. For example, if someone were to ask me why I think there is ``a 75\% chance of rain this afternoon", I could tell you that from my experience I have come to believe that here in New England there is a 50\% prior chance of rain on any given day, and moreover that rain is often portended by grey clouds. Using Bayes' formula I can take into account the datum that I saw grey clouds in the sky this morning, which leads me to revise my probability for afternoon rain upwards to 75\%. Although this is more detail than they probably wanted to hear, it would certainly satisfy the questioner.

While QBism is therefore compatible with interpreting probability assignments in terms of the ``reasons" that ``justify" them, it certainly downplays this interpretation. This is because, as mentioned in Sec.\ \ref{sec:SubjectObject}, QBism places minimal restrictions on what agents' prior beliefs can be: any arbitrary probability assignment is assumed to be \tit{ab initio} justifiable by some (unspecified) reasons. Put another way, the QBist tacitly assumes that \tit{reasons can always be contrived to justify any initial prior}. Under this assumption, one is not interested in finding out why agents believe what they do; rather, one is interested in calling upon the agents to act upon their beliefs, say, by having them place bets with meaningful stakes and seeing how they fare.

Bringing QBism in contact with Husserl's conception of probabilities therefore suggests that we should adopt a more balanced view that gives equal emphasis both to the reasons behind probability assignments (as captured by agents' priors and past history) and the future actions that the agents commit themselves to in making their assignments.

Fortunately, such a balanced approach has already been articulated in detail by the contemporary pragmatist philosopher Robert Brandom. Within his framework of ``inferentialism", Brandom proposes that the fundamental conceptual elements are ``assertions", which are defined as ``essentially performances that can both serve as and stand in need of reasons" \citep{brandom}. Moreover, Brandom considers an assertion as akin to a ``move" made in the context of a ``game of giving and asking for reasons". The ``game" is comprised of a set of normative rules, to which all agents under consideration are assumed to adhere. As Brandom points out, the two-sided nature of an assertion gives rise to two kinds of normative rules, which Brandom names ``commitments" and ``entitlements". As \citet{brandom} explains:

\begin{quote}
[A]ssertional games must have rules of this sort: rules of \tit{consequential commitment}. Why? Because to be recognizable as assertional, a move must 
not be idle, it must make a difference, it must have consequences for what else it is appropriate to do, according to the rules of the 
game. [\dots] Understanding a claim, the significance of an assertional move, requires understanding at least some of its consequences, knowing what else (what other moves) one would be committing oneself to by making that claim. [\dots] For this reason we can understand making a claim as taking up a particular sort of normative stance toward an inferentially articulated content. It is endorsing it, taking responsibility for it, committing oneself to it. --- \citet{brandom}, p191-192.
\end{quote}

And:

\begin{quote}
Giving reasons for a claim is producing other assertions that license or entitle one to it, that justify it. Asking for reasons for a claim is asking for its warrant, for what entitles one to that commitment. Such a practice presupposes a distinction between assertional commitments to which one is entitled and those to which one is not entitled. Reason-giving practices make sense only if there can be an issue as to whether or not practitioners are entitled to their commitments. Indeed, I take it that liability to demands for justification -- that is, demonstration of entitlement -- is another major dimension of the responsibility one undertakes, the commitment one makes, in asserting something. --- \citet{brandom}, p193.
\end{quote}

From these excerpts one can clearly see that what counts as ``good reasons" for an assertion, and what counts as ``consequent" upon an assertion, are established by normative rules, which are the ``entitlements" and ``commitments", respectively, that govern a specific ``game of giving and asking for reasons". By considering the various ways that these normative rules relate to one another, Brandom develops what he calls the ``normative fine-structure of rationality" (\citet{brandom}). 

One is immediately struck by the idea that a probability assignment may be interpreted precisely as a special kind of ``assertion" in Brandom's sense, whereby QBism's apparent divergence from Husserl on his second point may be understood as due to a difference in emphasis: Husserl's emphasis on an agent's ``entitlements" to a probability assertion on the one hand, and QBism's emphasis on an agent's ``commitments" in making a probability assertion; on Brandom's account these are two sides of the same coin. Furthermore, QBism so far only deals with the issue of how an agent's probability assignments  \tit{commit} them to certain actions (like buying lottery tickets); Brandom's concept of entitlement could enrich QBism by widening its scope to include how an agent's beliefs may or may not \tit{entitle} them to probability assignments of a certain form\footnote{There may already be a precedent for this idea in some QBist work. For instance, one could read the quantum de Finetti theorem (\cite{CFS_deFinetti}) as proving that if an agent believes that a set of systems is `exchangeable', they are thereby entitled to make a probability assignment that has a particular mathematical form.}.

The possible relevance of Brandom's work to quantum mechanics has been recognized previously in the work of Richard Healey \citep{healey}. The key difference between our present proposal and Healey's approach can be traced to two fundamentally different methodological starting points regarding the appropriate `set of possible agents'. As already discussed, QBism locates the quantum state at the `subjective' end of the spectrum of probabilistic assertions, namely, those which are assumed to differ \tit{a priori} among the given set of possible agents. By contrast, Healey, chooses to locate the quantum state in the `normative' component of quantum theory, assuming that there is a single quantum state that represents a norm of `best advice' for the given set of agents. Thus, on Healey's account, the quantum state is a set of probabilistic assertions that is `objective' in the sense of being invariant across the relevant set of agents. There is still much work to be done in comparing and contrasting different pragmatist approaches to quantum theory; an overview can be found in \citet{healey_sep}.

Inferentialism therefore not only can provide the missing piece that is needed to combine Husserl's phenomenology of probabilities with QBism's subjective decision-theoretic approach, but might also provide a unifying framework in which to compare and contrast QBism with other `pragmatic' interpretations of quantum theory including that of Healey. A fuller investigation of these themes must be left to future work.

\section{Conclusion}
We conclude by summarizing our findings. QBism and phenomenology have many points in common, and indeed a common goal: to articulate a metaphysics of ``experienced outcomes" or ``perceived phenomena" which are neither figments of the observer's mind, nor phantoms in a world beyond all observation. However it is by no means obvious that phenomenologists and QBists see eye to eye on every issue. We have examined the question of the existence of unobservable entities in science from both points of view, and compared their approaches. We found that QBism distinguishes between system-like entities and probability-like entities, and adheres to the Prolongation Thesis, which renders unobservable system-like entities potentially observable, contrary to Wiltsche's objection \citep{wiltsche_2012}. On the other hand, in order to maintain its claim that probability-like entities cannot be said to physically exist, QBism should embrace Wiltsche's ``Originality Thesis of Justification" over Berghofer's counter-proposed ``Criterion of Justification", as the latter opens the door to the possible physical existence of quantum states and hidden variables. 

Our rejection of Berghofer's criterion raised the question of what exactly are probability-like entities, if not things that physically exist. Here we encountered the difficulty that phenomenologists have said very little about the interpretation of probability. We argued that Husserl's account of probabilities as ``qualities of judgement" and ``weighted presumptions" could be made compatible with QBism if we interpret Husserl's usage of the word ``objective" to describe features of probability theory as being equivalent to QBism's usage of the word ``normative" to describe the invariant nature of these rules among rational agents. 

Furthermore we pointed out that although Husserl and QBism provide seemingly opposing accounts of the origin of the probability values, with Husserl appealing to the reasons that ``speak for" a probability assignment, and QBism appealing to the gambling commitments that stem from a probability assignment, a reconciliation can be found in Brandom's inferentialism, if we regard probabilities as assertions which entail both supporting reasons and consequent commitments. This leads us to interpret probability-like entities as collections of assertions made by an agent, which stem from the agent's prior probabilities and past experience, and serve to co-ordinate their future actions, within a context of normative rules. In this manner it may be possible to make sense of QBism's distinction between system-like and probability-like unobservable entities, and explain QBism's claim that system-like entities can be observable in principle (with an appropriate \nohyphens{intracorporeal} measuring instrument), while probability-like entities such as quantum states cannot.

\tit{Acknowledgments:} This work was supported in part by the John E. Fetzer Memorial Trust.

%\bibliographystyle{apalike}
%\bibliography{Unobservable}

\end{document}